\documentclass[twocolumn,amsmath,amssymb,prapplied,longbibliography]{revtex4-2}

\usepackage{graphicx}% Include figure files
\usepackage{dcolumn}% Align table columns on decimal point
\usepackage{bm}% bold math
\usepackage{color}
\usepackage{notes2bib}
\usepackage{soul}

\begin{document}

\bibliographystyle{naturemag}

%\preprint{}

\title{Telecom-band quantum memory with chlorine defects in silicon carbide}
%\title{Telecom spin-photon interface enabled by chlorine defects in silicon carbide}

\author{A.~N.~Anisimov$^{1}$}
\email[E-mail:~]{a.anisimov@hzdr.de}
\author{K.~Mavridou$^{1}$}
\author{A.~V.~Mathews$^{1,2}$}
\author{M.~Helm$^{1,2}$}
\author{G.~V.~Astakhov$^{1}$}
\email[E-mail:~]{g.astakhov@hzdr.de}

\affiliation{$^1$Helmholtz-Zentrum Dresden-Rossendorf, Institute of Ion Beam Physics and Materials Research, 01328 Dresden, Germany  \\
$^2$Technische Universit\"at Dresden, 01062 Dresden, Germany
 }

\begin{abstract}
Realization of quantum memory with a photonic interface in the telecommunication bands in a wafer-scalable platform is a central requirement for long-distance quantum networks. Silicon carbide (SiC) provides a technologically mature host for integrated quantum photonics, yet only a limited number of defects combine spin functionality with telecom emission. Here we report on chlorine-based defects in 4H-SiC as a platform for telecom-band quantum memory. The emission of these defects spans the entire telecommunication range with zero-phonon lines in the O- and C-bands and a Debye-Waller factor of up to 39\%. Time-resolved photoluminescence measurements reveal a short excited-state lifetime in the sub-nanosecond range. We demonstrate that these defects are spin-active even at room temperature, exhibiting optically detected magnetic resonances (ODMR) in the sub-GHz frequency range. Using ODMR spectroscopy and Ramsey interferometry, we resolve the hyperfine structure arising from the interaction with $^{35}\mathrm{Cl}$ nuclear spins. The ODMR spectra exhibit complex behaviour in an external magnetic field due to mixing of electron-nuclear spin states, which is well reproduced by our simulations. The spin relaxation and coherence times are in the sub-microsecond range, limited by rapid quenching of the ODMR contrast and attributed to charge-state metastability. The combination of telecom-band emission, coherent spin control and compatibility with wafer-scale fabrication positions Cl-related defects in SiC as a promising platform for chip-scale quantum memories with spin-photon interfaces operating in the fiber-optic telecommunication windows.
\end{abstract}
 
\date{\today}

\maketitle
%---------------------------------------------------------------

%\section*{Introduction} 

Networked quantum systems 
%, composed of spatially separated quantum nodes interconnected by photonic channels, 
offer a scalable framework \cite{10.1038/nature07127} for secure quantum communication \cite{10.1126/science.aam9288}, distributed quantum computation \cite{10.1103/physreva.89.022317} and remote quantum sensing \cite{10.1002/lpor.201900097}. Central to such architectures are long-lived quantum memories, which enable the storage and coherent manipulation of quantum information \cite{10.1103/physreva.71.060310}. Solid-state implementations based on localized atomic spins are particularly promising in this context due to their long coherence times and compatibility with scalable material platforms \cite{10.1103/physrevx.4.031022, 10.1063/5.0049372, 10.1103/prxquantum.5.010102, 10.1063/5.0262377, 10.1063/5.0056534, 10.1103/prxquantum.1.020102}. From a materials perspective, a crucial requirement is a well-established and efficient interface between spin-based quantum memories and photons, enabling their integration into quantum optoelectronic devices \cite{10.1038/nature15759, 10.1126/science.aax9406, 10.1038/s41586-020-2441-3, 10.1038/s41563-021-01148-3, 10.1038/s41586-023-06281-4, 10.1038/s41566-025-01752-8}. Such a spin-photon interface, operating in the low-loss spectral windows of fiber-optic telecommunications, enables reliable and long-distance distribution of quantum states. Despite this promise, the simultaneous realization of long-lived spin coherence and a native, efficient spin-photon interface in the telecom bands remains an outstanding materials challenge, making this combination a key bottleneck for scalable quantum networks.

Prominent examples of spin-based quantum memories with spin-photon interfaces are nitrogen-vacancy (NV) \cite{10.1126/science.abg1919}, silicon-vacancy (SiV) \cite{10.1126/science.add9771} and tin-vacancy (SnV) \cite{10.1038/s41586-024-07371-7} defects in diamond, which have enabled landmark experimental demonstrations of photon-mediated remote entanglement between spin qubits. However, the optical transitions of these color centers lie in the visible spectral range, necessitating frequency conversion to the telecom bands for fiber-based networking  \cite{10.1103/PhysRevLett.123.063601}. Such conversion typically requires additional active and passive optical components, increasing system complexity and posing challenges for scalable quantum network architectures. Color centers in silicon represent an alternative platform for the implementation of telecom-band quantum memories with spin-photon interfaces owing to their intrinsic compatibility with complementary metal-oxide-semiconductor (CMOS) fabrication protocols and emission in the telecom bands \cite{10.1038/s41467-022-35051-5}. To date, such implementations have been realized primarily using T centers \cite{10.1038/s41586-022-04821-y} and erbium dopants \cite{10.1038/s41467-024-55552-9}. However, their practical deployment is constrained by significantly lower photon emission rates compared with color centers in diamond, arising from intrinsically long radiative lifetimes.

In this context, silicon carbide (SiC) emerges as a promising host material that bridges these limitations by combining bright color centers with wafer-scale availability and fabrication processes compatible with established CMOS technologies \cite{10.1063/5.0004454, 10.1063/5.0262377}. Among the various defect centers investigated in SiC for spin-photon quantum technologies, silicon vacancies ($\mathrm{V_{Si}}$) \cite{10.1103/physrevlett.109.226402} and divacancies \cite{10.1038/nature10562} have attracted particular attention owing to their spin-dependent optical transitions \cite{10.1038/nphys2826, 10.1126/sciadv.1501015}, long coherence times  \cite{10.1103/physrevb.95.161201} and suitability for single-defect control \cite{10.1038/nmat4145, 10.1038/nmat4144, 10.1038/ncomms8578, 10.1038/s41467-019-09873-9}.  A key limitation of the $\mathrm{V_{Si}}$ and divacancy defects in SiC is that their optical emission lies in the near-infrared spectral range, such that wavelength conversion into the telecom bands is still required. %, as in the case of diamond color centers. 

Alternatively, vanadium-related defects in SiC, which feature an established spin-photon interface operating in the telecom O-band, have attracted considerable attention in spite of their comparatively low emission brightness and complexity of spin  control required sub-Kelvin temperatures \cite{10.1126/sciadv.aaz1192, 10.1038/s41467-023-43923-7, 10.1103/physrevapplied.22.044078}. Therefore, the realization of an NV- or divacancy-like centers in SiC that combines robust long-lived spin coherence with a native spin-photon interface in the telecom bands would represent a highly desirable advance, with the potential to fundamentally change the design paradigm of solid-state quantum networks and constitute a significant step toward their practical implementation. 

It has been theoretically predicted \cite{10.1103/physrevb.108.224106}  and recently experimentally demonstrated \cite{10.1364/oe.581705} that chlorine-vacancy (ClV) defects in SiC emit in the telecom O- and C-bands, which are the most widely used spectral windows for data-center networking and long-distance fiber-optic communication. A  wavefunction theory anyalysis suggests the existence of an optical spin polarization cycle for the ClV defects \cite{10.48550/arxiv.2511.21965}. However, this mechanism has not yet been experimentally verified and coherent spin control, which an essential prerequisite for quantum memory functionality, has so far not been demonstrated.

Here we report on chlorine-related defects in 4H-SiC that combine telecom-band emission with coherent spin control in a wafer-scalable material platform. These defects are created using a standard ion implantation procedure in commercially available SiC wafers. They exhibit zero-phonon lines (ZPLs) in the telecom O- and C-bands. Power-dependent excitation reveals additional ZPLs at shorter wavelengths, which we attribute to excited states (ES) of the Cl-related defects. We measure a Debye-Waller factor of up to 39\%, significantly exceeding that of NV centers in diamond and intrinsic vacancy related defects in SiC. Time-resolved photoluminescence (PL) measurements yield an ES lifetime of $450\,\mathrm{ps}$, shorter than typically reported for color centers. 

Furthermore, we demonstrate a ground-state (GS) spin-photon interface, with zero-field-splitting parameters giving rise to ODMR resonances in the range of $500 - 700 \,\mathrm{MHz}$ and a hyperfine interaction constant $|  A  |  = 34 \,\mathrm{MHz}$. In addition, we observe ORMR resonances in the range of $100 - 400 \,\mathrm{MHz}$, which have likewise not been reported previously. These resonances are not related to Cl-related defects, but are instead associated with unidentified intrinsic optically active defects emitting in the telecom band. Coherent manipulation of the Cl-related defect spin is achieved using pulsed radiofrequency (RF) control. The observed Ramsey fringes reveal hyperfine interaction with $^{35}\mathrm{Cl}$ nuclear spins and spin coherence on sub-microsecond timescales. We show that this coherence time is not intrinsic to the chlorine-related defects, but is likely limited by charge-state dynamics. Mixing of electron-nuclear spin states in an external magnetic field gives rise to a dense manifold of closely spaced transitions, which is reproduced in good agreement with our simulations. These results establish Cl-related defects in SiC as a promising platform for chip-scale quantum memories in long-distance quantum networks.

\section{Spin-photon interface}

\begin{figure*}[t]
\includegraphics[width=.95\textwidth]{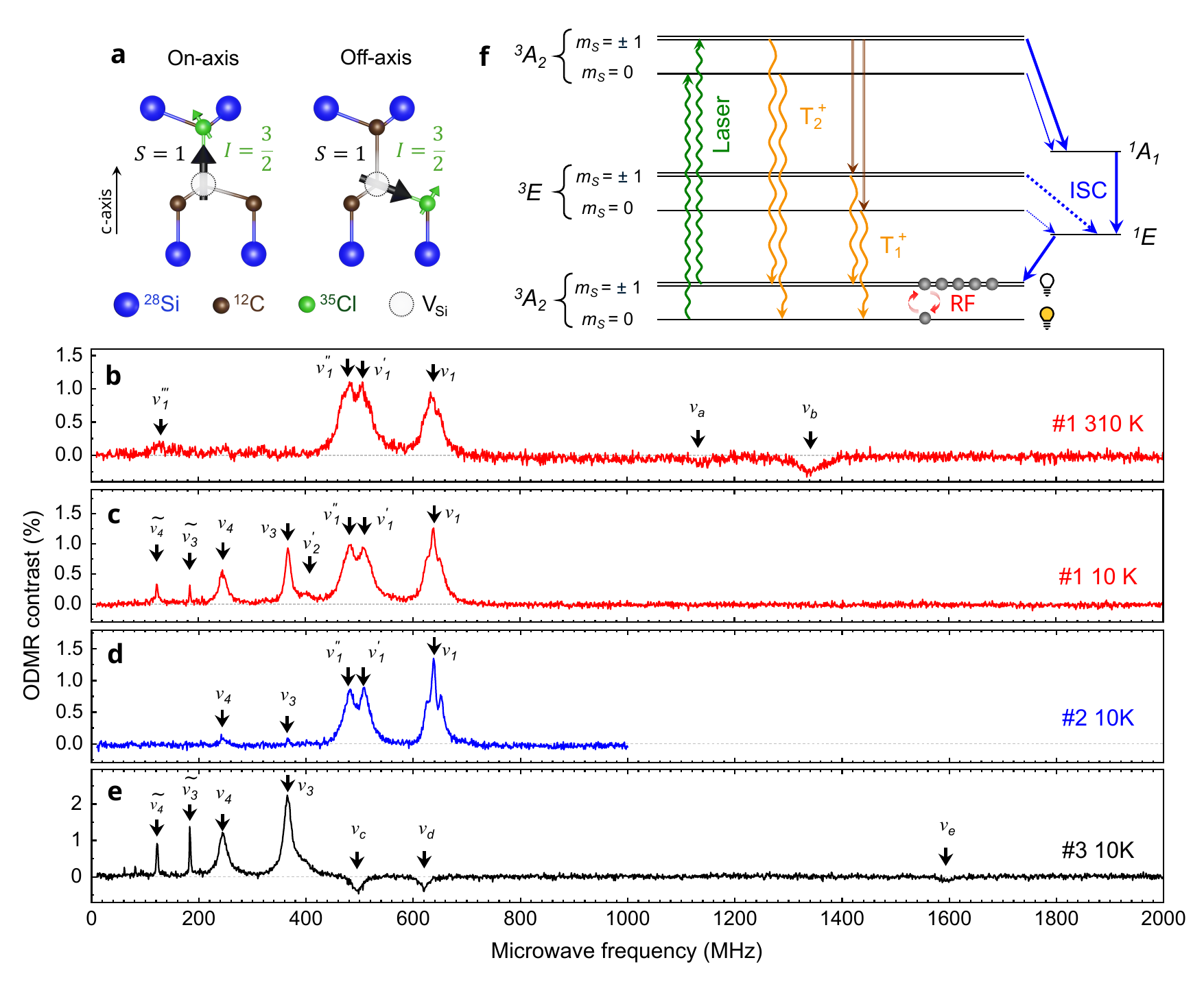}
\caption{ODMR of Cl-related defects in 4H-SiC.
(a) Schematic illustration of on-axis and off-axis ClV defects with electron spin $S = 1$ coupled to a $^{35}\mathrm{Cl}$ nuclear spin $I = 3/2$.
(b) Room-temperature ODMR spectrum measured in a $^{35}\mathrm{Cl}$-implanted 4H-SiC wafer (sample $\#1$) under optical excitation at $\lambda_{\mathrm{ex}} = 976 \,\mathrm{nm}$ and PL detection at $\lambda_{\mathrm{PL}} > 1300 \,\mathrm{nm}$. Arrows indicate the spin resonances discussed in the main text.
(c) Same as in (b), measured at low temperature ($T = 10 \,\mathrm{K}$) for sample $\#1$. (d) Same as (c) measured at low temperature ($T = 10 \,\mathrm{K}$) for sample $\#2$. (e) Same as (c) measured at low temperature ($T = 10 \,\mathrm{K}$) for sample $\#3$. (f) Schematic representation of the optical spin polarization cycle of the ClV defect,  involving a second ES.
}\label{fig1}
\end{figure*}

According to the DFT calculations  \cite{10.1103/physrevb.108.224106}, the ClV consist of a chlorine atom substituting a carbon atom in the SiC lattice, paired with an adjacent silicon vacancy (Fig.~\ref{fig1}a). In 4H-SiC polytype, there are two on-axis configurations with symmetry $C_{3v}$, when the defect axis is parallel to  c-axis of the crystal, and two off-axis configurations with lower symmetry $C_{1h}$. The telecom-active  chlorine-vacancy defects are predicted to exist in a positively charged state $\mathrm{ClV^+}$ and to possess a non-zero electron spin $S = 1$ \cite{10.1103/physrevb.108.224106}, analogous to divacancies in 4H-SiC and NV centers in diamond. In contrast, the neutral charge state  $\mathrm{ClV^0}$ is predicted to be stable with spin $S = 1/2$, which does not support zero-field ODMR.

In our experiments reported here, we concentrate on a weak p-type epitaxially grown layer with a thickness of $250 \, \mathrm{nm}$ (sample $\#1$) and a weak n-type epitaxially grown layer with a thickness of $20 \, \mathrm{\mu m}$ (sample $\#2$).  The Cl-related defects were created by implantation of $^{35}\mathrm{Cl}$ ions followed by thermal annealing, using the protocol described in our previous work \cite{10.1364/oe.581705}.  The $^{35}\mathrm{Cl}$ isotope possesses a nuclear spin of $I = 3/2$, resulting in a coupled electron-nuclear spin system (Fig.~\ref{fig1}a).  As a reference, the same wafer as for sample $\#1$ is used, but implanted with $^{40}\mathrm{Ar}$ instead of $^{35}\mathrm{Cl}$ under identical implantation energy, fluence, and thermal treatment conditions (sample $\#3$). Although our experiments do not allow us to unambiguously identify the observed PL and ODMR signatures with a specific defect configuration, the data are qualitatively consistent with the ClV model, despite some quantitative discrepancies. While alternative configurations cannot be excluded, the ClV center currently represents the only theoretically established Cl-related defect in SiC that combines telecom emission with spin activity. We therefore base our interpretation of the experimental results on this framework \cite{10.1103/physrevb.108.224106, 10.48550/arxiv.2511.21965}.
 
To demonstrate the robustness of the spin-photon interface of the ClV defects, we begin with ODMR measurements at room temperature and in zero magnetic field. The ClV PL is excited using a $976 \, \mathrm{nm}$ laser and detected through a longpass filter with a cutoff wavelength of $1300 \, \mathrm{nm}$, corresponding to the spectral region of ClV emission. Figure~\ref{fig1}b shows the resulting ODMR spectrum obtained in sample $\#1$, measured over an RF frequency range from $10 \, \mathrm{MHz}$ to $2 \, \mathrm{GHz}$. We first note that the ODMR resonances of divacancies and NV-related defects in 4H-SiC, whose phonon sideband (PSB) may exhibit a decaying tail extending to wavelengths $\lambda_{\mathrm{PL}} > 1300 \,\mathrm{nm}$, typically occur in the RF frequency range of $1.1 - 1.4 \,\mathrm{GHz}$. Broad and weak ODMR feature $\nu_a$ and $\nu_b$ of unidentified origin could therefore, in principle, be attributed to the PL4 divacancy or the NV (hh) defect. However, its substantial linewidth of approximately $40\,\mathrm{MHz}$ instead can alternatively suggest a potential origin in an ES of the ClV defects, which provides the dominant contribution to the PL. In the low-frequency spectral range below $700 \,\mathrm{MHz}$, multiple ODMR resonances labeled $\nu_1$, $\nu_1'$, $\nu_1''$ and $\nu_1'''$ are observed in Fig.~\ref{fig1}b. Although divacancies and NV-related defects in 4H-SiC are detectable in this sample, their PL intensity is significantly lower than that of the ClV defects \cite{10.1364/oe.581705}, and they do not exhibit identified resonances in this RF frequency range. We therefore tentatively attribute these features to spin transitions within the GS multiplet of one of the ClV defect, as discussed in detail below.

At low temperature $T = 10 \, \textrm{K}$, in addition to $\nu_1 = 640 \,\mathrm{MHz}$, $\nu_1' = 509 \,\mathrm{MHz}$, $\nu_1'' = 482 \,\mathrm{MHz}$ and $\nu_1''' = 122 \,\mathrm{MHz}$,  two further resonances $\nu_{3} = 367 \,\mathrm{MHz}$ and $\nu_{4} = 245 \,\mathrm{MHz}$ with comparable ODMR contrast are observed sample $\#1$ (Fig.~\ref{fig1}c). From the magnetic-field dependence presented below, we attribute these features to defects with hyperfine interactions distinct from those of the $\nu_1$-resonance family, for which the hyperfine structure is resolved, whereas it remains unresolved for the present features. Apart from that, two much narrower ODMR resonances $\widetilde{\nu}_{3} = 184 \,\mathrm{MHz}  \approx  \nu_{3}  / 2$ and $\widetilde{\nu}_{4} = 122 \,\mathrm{MHz} \approx  \nu_{4} / 2$ can be attributed to two-photon spin transitions. As such processes are generally weak, we cannot exclude the possibility that these resonances are enhanced by hyperfine coupling between the electron and nuclear spins, potentially mediated by mutual spin flip-flop processes. Finally, we also observe a weak resonance at $\nu_2' = 406 \,\mathrm{MHz}$. A detailed analysis presented below further reveals the presence of additional resonances, $\nu_2$ and $\nu_2''$. Their spectral behavior closely resembles that of the $\nu_1$-resonance family. 

In a second Cl-implanted wafer (sample $\#2$), the low-temperature ODMR spectrum is dominated by the $\nu_1$-resonance family, with only weak signatures of the $\nu_3$ and $\nu_4$ resonances (Fig.~\ref{fig1}d). In contrast, in the Ar-implanted wafer (sample $\#3$), the $\nu_1$ resonance family is absent, while the $\nu_3$ and $\nu_4$ resonances are clearly observed (Fig.~\ref{fig1}e). In addition, ODMR lines $\nu_c$, $\nu_d$, and $\nu_e$ are detected only in the Ar-implanted wafer and are not observed in the Cl-implanted wafers. Based on the ODMR spectra in Fig.~\ref{fig1}b-e, we attribute the $\nu_1$ and weak $\nu_2$ resonances to Cl-related defects, whereas the $\nu_3$ and $\nu_4$ resonances appear to be wafer-specific. The additional resonances $\nu_{a-e}$  remain of unidentified origin and are not investigated further in this work.

%We thus ascribe the four observed ODMR resonance families to distinct crystallographic configurations of the ClV defect, comprising two on-axis and two off-axis orientations (Fig.~\ref{fig1}a). However, this assignment is not straightforward and requires a detailed analysis of the correlation between the ODMR resonances and the PL spectrum. 

A schematic of the optical transitions and the spin polarization cycle in $\mathrm{ClV^+}$ is shown in Fig.~\ref{fig1}f. The scheme is adapted from theoretical calculations of the transition rates and intersystem crossing processes \cite{10.48550/arxiv.2511.21965}. Owing to the crystal field, the electron spin ($S = 1$) in the $^3A_2$ GS and the $^3E$ ES is split into sublevels with spin projections $m_S = 0$ and $m_S = \pm 1$ on the defect axis. In addition, singlet states ($S = 0$) are also present.  One of them, $^1E$, lies energetically between the $^3A_2$ GS and the $^3E$ ES, while another singlet state, $^1A_1$,  is located above the $^3A_2$ GS. The intersystem crossing (ISC) transitions from the $m_S = \pm 1$ sublevels of the $^3E$ ES to the $^1E$ state, and from $^1E$ back to the $m_S = \pm 1$ sublevels of the $^3A_2$ GS, are predicted to be significantly faster than those involving the $m_S= 0$ sublevel. This results in lower emission probability for $m_S = \pm 1$, which is the dark spin state. The observation of a positive ODMR contrast under RF driving indicates that the $m_S= \pm 1$ sublevels of the $^3A_2$ GS are preferentially populated via the ISC process. This behavior further implies that the transition from the $m_S = 0$ sublevel of the $^3E$ ES to the $^1E$ state is faster than the reverse transition from $^1E$ to the $m_S = 0$ sublevel of the $^3A_2$ GS. For clarity, the latter transition is omitted in Fig.~\ref{fig1}f.

\begin{figure*}[t]
\includegraphics[width=.95\textwidth]{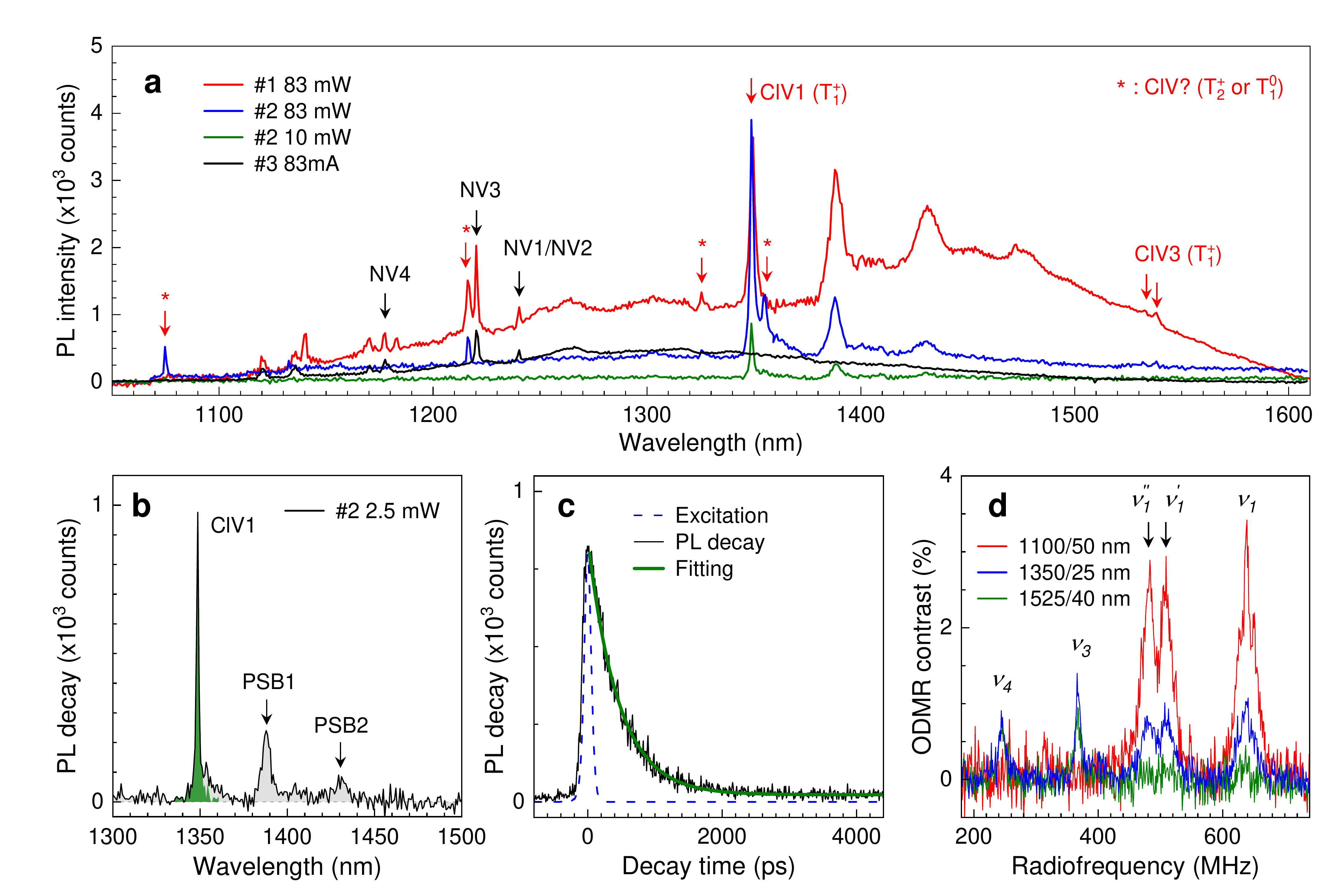}
\caption{Optical spectroscopy and photoexcitation dynamics of Cl-related defects in 4H-SiC. (a) PL spectra from different samples recorded at different excitation powers with $\lambda_{\mathrm{ex}} = 976 \,\mathrm{nm}$. (b) PL recorded with an excitation power of $2.5 \,\mathrm{mW}$ in sample $\#2$, yielding the Debye-Waller factor of 39\%. (c) Time-resolved PL measurement (thin solid line) under pulsed excitation at $\lambda_{\mathrm{ex}} = 1060 \,\mathrm{nm}$ and pulse duration of $10 \, \mathrm{ps}$ (dashed line) in sample $\#1$.  The ClV1 ZPL is spectrally selected with a bandpass filter with a central wavelength of $1350 \, \mathrm{nm}$ and a bandwidth of $12 \, \mathrm{nm}$ in sample $\#1$. A fit using a convolution of the laser pulse profile with a mono-exponential decay yields the ES life time of  $450 \pm 10 \, \mathrm{ps}$ (thick solid line).  (d) RF-assisted spectroscopy of Cl-related spins in sample $\#1$. ODMR spectra are recorded using spectrally distinct detection windows.} 
\label{fig2}
\end{figure*}

The optical transition between the $^3E$ ES and the $^3A_2$ GS is denoted as $\mathrm{T}_1^+$ in recent DFT calculations \cite{10.1103/physrevb.108.224106}. This transition is optically allowed and gives rise to the zero-phonon line (ZPL) in the PL spectrum. The ClV1 ZPL dominates the PL spectrum under $976 \,\mathrm{nm}$ excitation at $T = 10 \,\mathrm{K}$ across different Cl-implanted  wafers (samples $\#1$ and $\#2$) and absent in the Ar-implanted wafer (sample $\#3$), as shown in Fig.~\ref{fig2}a. We therefore associate this ZPL at $\lambda_{\mathrm{ZPL}} = 1348.7 \, \mathrm{nm}$ with the $\mathrm{T}_1^+$ transition in the telecom O-band. In previous work, this transition was attributed to the on-axis configuration of the ClV defect \cite{10.1364/oe.581705}. 
 At a low excitation power ($2.5 \,\mathrm{mW}$), only the ClV1 ZPL accompanied by two peaks at $\lambda_{\mathrm{PSB1}} = 1387.9 \, \mathrm{nm}$ and $\lambda_{\mathrm{PSB2}} = 1431.1 \, \mathrm{nm}$  in the phonon sideband (PSB) are observed (Fig.~\ref{fig2}b). These features correspond to a local vibrational mode energy of approximately $25 \, \mathrm{meV}$. From integration of the PL spectrum over the range  $1300 - 1500 \, \mathrm{nm}$, we extract the Debye-Waller factor of 39\%. This value is significantly larger than that of NV centers in diamond of about 3\% \cite{10.1088/1367-2630/16/7/073026} or $\mathrm{V_{Si}}$ centers and divacancies in SiC of about 6\% \cite{10.1038/nmat4144, 10.1103/physrevb.101.144109}.

Figure~\ref{fig2}c shows a PL decay of the ClV1 ZPL under pulsed excitation, which is spectrally selected with a 1350/12 bandpass filter (central wavelength $1350 \, \mathrm{nm}$, bandwidth $12 \, \mathrm{nm}$). This notation is adopted throughout to denote all bandpass filters for brevity. By fitting the data using a convolution of the laser pulse profile with a mono-exponential decay function, we extract an ES lifetime  of $450 \pm 10 \, \mathrm{ps}$. This value is by two orders of magnitude shorter than the theoretically calculated lifetime \cite{10.48550/arxiv.2511.21965}. We emphasize that it does not necessarily reflect the intrinsic radiative recombination time, as our experimental conditions do not allow us to exclude contributions from nonradiative recombination channels, which may result from incomplete lattice recovery after implantation and annealing. %Time-resolved PL spectroscopy of the ClV defects at room temperature reveals a broad emission spectrum without a discernible ZPL and an even shorter ES lifetime $\tau = 265 \pm 3 \, \mathrm{ps}$ (Supplementary Figure~S2). 

At higher excitation powers, additional ZPLs in the telecom C-band associated with ClV3 are observed (Fig.~\ref{fig2}a), consistent with previous experimental reports and attributed to the off-axis configuration \cite{10.1364/oe.581705}. In addition, four ZPLs labeled with asterisks (*) in Fig.~\ref{fig2}a, emerge in the short-wavelength region of the PL spectrum at $1075 \, \mathrm{nm}$,  $1216 \, \mathrm{nm}$, $1326 \, \mathrm{nm}$ and $1355 \, \mathrm{nm}$. These ZPLs are distinct from the well-established NV ZPLs in 4H-SiC \cite{10.1103/physrevb.94.060102}. One possible interpretation is that they originate from the $\mathrm{T}_1^0$ transition, which is predicted for the neutral charge state $\mathrm{ClV^0}$ to occur at higher transition energies  \cite{10.1103/physrevb.108.224106}. However, $\mathrm{ClV^0}$ is not spin active, in contradiction with the ODMR measurements presented below. We therefore preliminary associate these ZPLs with the $\mathrm{T}_2^+$ optical transition from a second ES, as schematically depicted in Fig.~\ref{fig1}d and also predicted theoretically \cite{10.1103/physrevb.108.224106}.

To elucidate the origin of the ODMR resonances we  performed RF-assisted spectroscopy \cite{10.1103/physrevb.101.144109}. Using tunable bandpass filters, we select different spectral windows of the PL spectrum and perform ODMR measurements. The ODMR spectrum detected in the vicinity of the ClV1 ZPL using a tunable 1350/25 bandpass filter is presented in Fig.~\ref{fig2}d. Comparison with Fig.~\ref{fig1}c demonstrates that spectral filtering does not alter the set of observed ODMR resonances, although their relative amplitudes are modified.  When a 1525/40 bandpass filter is used for detection, the ODMR spectrum is dominated by the resonances $\nu_3$ and $\nu_4$. In contrast, detection in the shorter-wavelength region of the PL spectrum using a 1100/50 bandpass filter reveals a dominant contribution from the $\nu_1$, $\nu_1'$, and $\nu_1''$ resonances. The relative amplitudes of these latter resonances remain nearly unchanged under different detection conditions, supporting their assignment to a common defect origin.

The change in PL intensity induced by the RF field tuned to the $\nu_1$ resonance reveals two distinct spectral regions, namely, a high response in the range $1200-1325 \,\mathrm{nm}$ and a lower response in the range $1350-1500\,\mathrm{nm}$. We interpret this behavior by the existing of a second ES, which is denoted in our scheme of Fig.~\ref{fig1}f as a higher-lying triplet state $^3A_2$.  For the $^3A_2$ ES, an additional ISC channel to the $^1A_1$ singlet state becomes available (Fig.~\ref{fig1}f). Recent theoretical work predicts that this channel has a higher ISC rate than the pathway through the $^1E$ singlet state \cite{10.48550/arxiv.2511.21965}, which may contribute to the enhanced ODMR contrast. 

%We tentatively assign the $\nu_1$ resonances to ClV1. The weak $\nu_2$ resonance can be assigned to ClV2. 

\section{Electron-nuclear spin system}

\begin{figure*}[t]
\includegraphics[width=.95\textwidth]{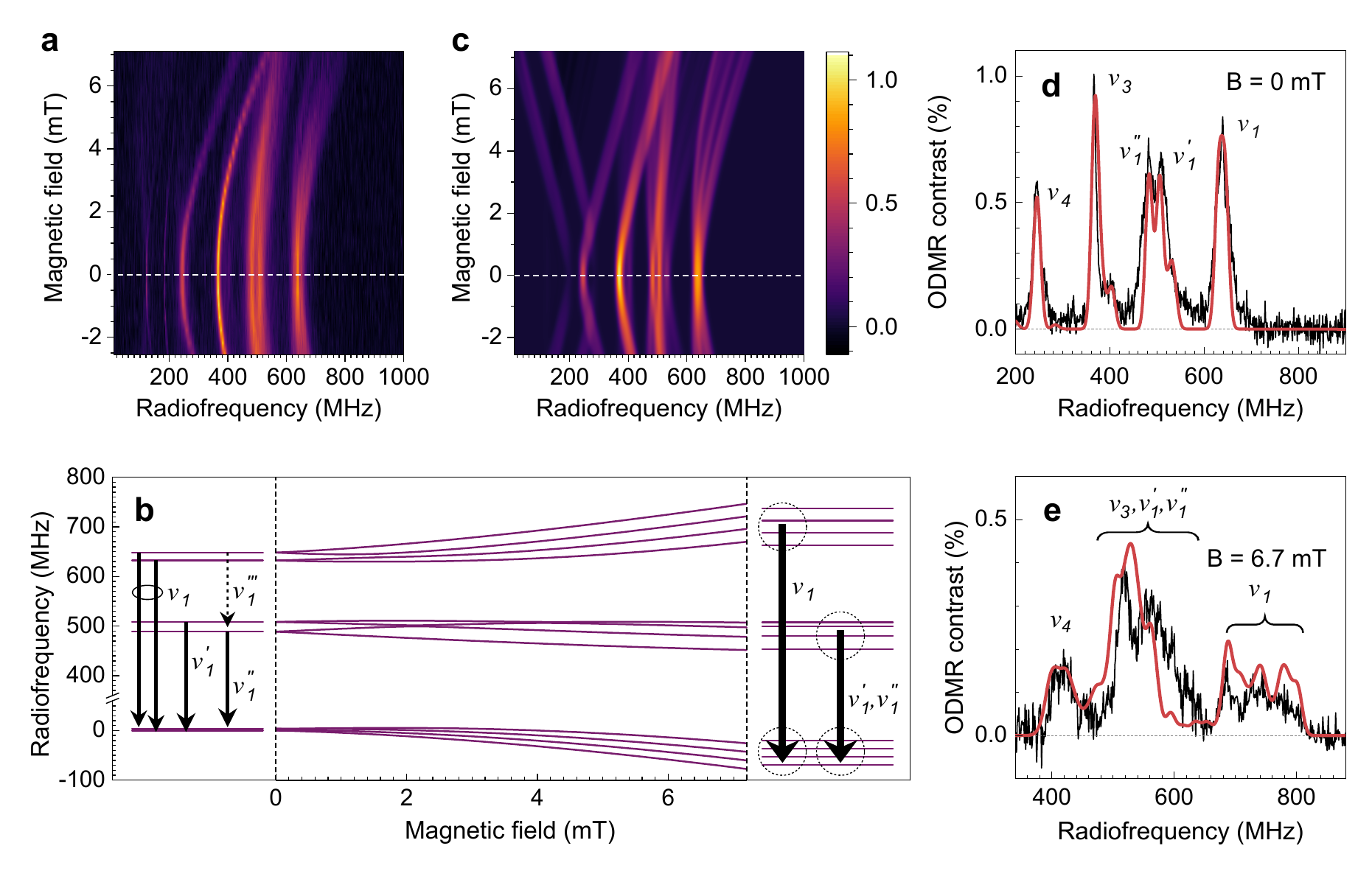}
\caption{Coupled electron-nuclear spins of Cl-related defects in 4H-SiC. (a) Evolution of the experimentally measured ODMR spectrum as a function of magnetic field applied along the c-axis in sample $\#1$. (b) Spin eigenstates as a function of the magnetic field calculated for the $\nu_1$ resonance family with parameters from table~\ref{tab:parameters}.  The arrows indicate transitions corresponding to the $\nu_1$, $\nu_1'$, $\nu_1''$ and $\nu_1'''$ resonances. (c) Evolution of the theoretically calculated ODMR spectrum as a function of magnetic field applied along the c-axis. (d) Comparison of the measured and calculated ODMR spectra in zero magnetic field used to obtain parameters of the spin Hamiltonian~(\ref{eq:H}). (e) Comparison of the measured and calculated ODMR spectra in magnetic field $B = 6.7 \, \mathrm{mT}$ without additional fitting parameters.}
\label{fig3}
\end{figure*}

The detection of RF-induced changes in the PL intensity alone does not unambiguously establish spin resonances as their origin. A key criterion is their dependence on an external magnetic field. Furthermore, to demonstrate the involvement of chlorine, it is essential to resolve the hyperfine structure arising from coupling to the chlorine nuclear spins. To address these points, we performed ODMR measurements as a function of magnetic field $B$ applied along the c-axis (Fig.~\ref{fig3}a). Indeed, we observe a clear shift of all spin resonances with increasing magnetic field.

%\begin{align}\label{eq:H}
%H = &D \left[S_z^2 - \frac{S(S+1)}{3}\right] + g\mu_B \bm B \cdot \bm S \nonumber\\
%&+ A_\parallel S_z I_z + A_\perp (S_xI_x+S_yI_y) ,
%\end{align}

%\begin{align}\label{eq:H}
%H = &D S_z^2  + E (S_x^2 - S_y^2) + \gamma_e \bm B \cdot \bm S \nonumber + \gamma_n \bm B \cdot \bm I  + \bm S \cdot \bm A  \cdot  \bm I  ,
%\end{align}

The spin Hamiltonian for the ClV defect in external magnetic field is given by
\begin{align}\label{eq:H}
H = H_{\mathrm{ZF}} +  H_{\mathrm{EZ}} + H_{\mathrm{NZ}} + H_{\mathrm{HF}} \,.
\end{align}
Here, the crystal field is described by the zero-field interaction $H_{\mathrm{ZF}}  = D (S_z^2 - S (S+1)/3)  + E (S_x^2 - S_y^2)$, where $D$ and $E$ are the zero-field splitting parameters parallel and perpendicular to the defect axis, respectively. The magnetic-field dependence is dominated by the electron Zeeman term $H_{\mathrm{EZ}}   =  \gamma_e \textbf{B} \cdot \textbf{S}$, with the gyromagnetic ratio $\gamma_e = \mu_B g_e = 28 \,\mathrm{MHz \, mT^{-1}}$, where $\mu_B$ is the Bohr magneton and $g_e$ is the electron g factor. 
The nuclear Zeeman term $H_{\mathrm{NZ}}   =  - \mu_n g_n \textbf{B} \cdot \textbf{I}$,
with $\mu_n$ being the nuclear magneton  and  $g_n $ being the $^{35}$Cl nuclear g factor, has negligible contrinbution. % of $^{35}$Cl $g_n = 0.55$
The hyperfine interaction is described by $H_{\mathrm{HF}}   =  \textbf{S} \cdot \textbf{A} \cdot \textbf{I}$,  where $\textbf{A}$ is the hyperfine tensor. In the simplest case, $\mathbf{A}$ is diagonal and can be characterized by an effective hyperfine coupling constant  $A$.  
We use MATLAB to solve the spin Hamiltonian~(\ref{eq:H}) and simulate the ODMR spectra. 

Figure~\ref{fig3}b shows the calculated eigenvalues for the $\nu_1$ resonance family using the spin Hamiltonian parameters summarized in table~\ref{tab:parameters}. In the absence of an external magnetic field $B = 0 \, \mathrm{mT}$, the ClV defect exhibits six doubly degenerate states, expressed in the $|m_S, m_I \rangle$ basis. The $\nu_1'$ and $\nu_1''$ ODMR lines originate from the transitions between states $|\pm 1, \pm 1/2\rangle  \rightarrow |0, \pm 1/2\rangle$ and $|\pm 1, \pm 3/2\rangle  \rightarrow |0, \pm 3/2\rangle$, respectively.  The splitting between them depends on the hyperfine interaction in a non-trivial manner \cite{10.1038/s43246-025-00840-0}. The $\nu_1$ line also originates from two transitions $|\mp 1, \pm 1/2\rangle  \rightarrow |0, \pm 1/2\rangle$ and $|\mp 1, \pm 3/2\rangle  \rightarrow |0, \pm 3/2\rangle$, which are not resolved in the ODMR spectrum.  The weak $\nu_1'''$ ODMR line originates from the transition  $|\pm 1, \pm 1/2\rangle  \rightarrow |\mp 1/2, \pm 1/2\rangle$ and  $|\pm 1, \pm 3/2\rangle  \rightarrow |\mp 1/2, \pm 3/2\rangle$, which are forbidden for pure states but become available due to different mixing with the lowest  $m_S = 0$ states.  

\begin{table}[b]
\centering
\caption{Zero-field splitting parameters and hyperfine interaction constants obtained from simulations of the experimental ODMR spectra. The signs of $D$ and $A$ are taken from theoretical calculations \cite{10.1103/physrevb.108.224106}. }
\label{tab:parameters}
\begin{tabular}{|c|c|c|c|c|}
\hline
Spin resonance & Orientation & $D$ (MHz) & $E$ (MHz) & $A$ (MHz) \\
\hline
$\nu_1$  & off-axis & $560$ & $60$ & $-34$ \\
\hline
 $\nu_2$   &off-axis  & $455$ & $60$ & $-30$  \\
\hline
$\nu_3$  & on-axis & $326$ & $40$ &  \\
\hline
$\nu_4$  & on-axis &  $221$ & $20$ &  \\
\hline
\end{tabular}
\end{table}

In external magnetic fields $B \neq 0 \, \mathrm{mT}$, the degeneracy is lifted, resulting in 12 distinct mixed spin states and multiple transitions (Fig.~\ref{fig3}b). Simulations of the magnetic-field evolution of the ODMR spectra (Fig.~\ref{fig3}c) indicate that the ClV defect associated with the $\nu_1$ resonance family adopts an off-axis configuration. This contrasts with the on-axis configuration inferred earlier from the PL measurements. Resolving this discrepancy will require further theoretical work and may point to the existence of additional configurations of Cl-related defects.

Similar to $\nu_1$, the $\nu_3$ and $\nu_4$ resonances should exhibits corresponding counterpart in zero magnetic field. However, their ODMR contrast can be significantly reduced due to state mixing. Assuming that they possess $S = 1$ spin and based on their magnetic-field dependence, we extract the zero-field splitting parameters and infer an on-axis configuration, while the hyperfine interaction remains unresolved.

Figure~\ref{fig3}c presents the calculated magnetic-field evolution of the ODMR spectrum for all four  spin centers, demonstrating very good agreement with the experimental data in Fig.~\ref{fig3}a. The parameters of the spin Hamiltonian used in these calculations are obtained from the best fit to the ODMR spectrum at $B = 0 \, \mathrm{mT}$ and summarized in Table~\ref{tab:parameters}.  A comparison between the calculated ODMR spectrum in external magnetic $B = 6.7\,\mathrm{mT}$ and the experimental data is shown in Fig.~\ref{fig3}e.  The ODMR exhibits a complex structure arising from multiple allowed transitions, which is well reproduced by our calculations.

\section{Coherent spin dynamics}

\begin{figure}[t]
\includegraphics[width=.49\textwidth]{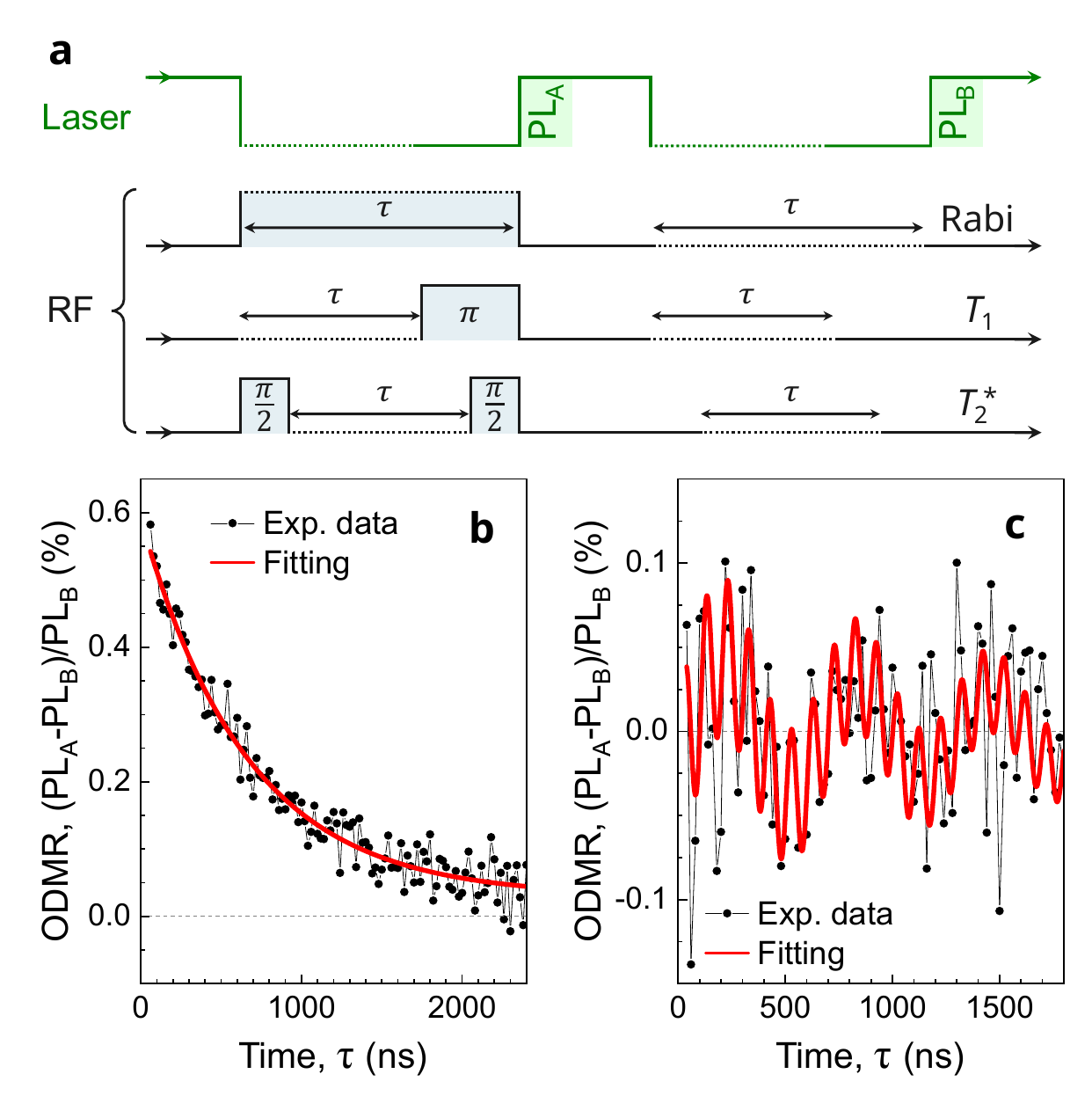}
\caption{Coherent spin dynamics of Cl-related defects in 4H-SiC. (a) The pulse sequence to perform Rabi, $T_1$ and $T_2^*$ measurements. (b) Measurement of spin lifetime at the $\nu_1$ resonance in zero magnetic field. The solid line is a mono-exponential fit yielding a decay constant of $660 \, \pm \, 35 \,\mathrm{ns} $. 
(c) Measurement of Ramsey fringes at the $\nu_1$ resonance arising from the hyperfine structure. The solid line shows a fit to Eq.~(\ref{eq:Ramsey}). % with $T_2^* = 2 \pm 1 \, \mathrm{\mu s}$.  
}\label{fig4}
\end{figure}

A critical requirement for the realization of solid-state quantum memories based on the ClV defects is the ability to prepare their spins in a coherent superposition and store this state for a finite time. As a first step, we investigate the spin-dependent PL dynamics upon switching the RF driving field on and off  when the RF is tuned to the $\nu_1$ resonance in nearly zero magnetic field. The PL intensity increases when the RF field is switched on, reflecting rapid averaging of the ClV spin populations between the $m_S = 0$ and $m_S = \pm 1$ states (Fig.~\ref{fig1}f). When the RF field is switched off, optical excitation initializes the ClV spin into the $m_S = \pm 1$ states. Because these states exhibit lower emission probability, the overall PL intensity decreases. From a mono-exponential fit, we extract a characteristic time of $300 \,\mathrm{ns}$ for the reinitialization of the ClV spins under optical pumping. Based on these measurements, we choose a PL detection window $t_{\mathrm{det}} = 250 \,\mathrm{ns}$, which is sufficiently long to ensure a strong PL signal while remaining short enough to avoid significant repumping of the spin state during detection.

For RF driving of the ClV spins, we employ a standard pulse sequence consisting of a $5\,\mathrm{\mu s}$ laser pulse for spin initialization, followed by an RF pulse of variable duration $\tau$, and a second laser pulse used for PL detection within a time window $t_{\mathrm{det}}$, during which the photon counts $\mathrm{PL_A}$ are recorded (Fig.~\ref{fig4}a). In a reference sequence, the RF pulse is omitted while all other timing parameters remain identical, yielding photon counts $\mathrm{PL_B}$. The ODMR contrast as a function of $\tau$ is then defined as $C = (\mathrm{PL_A}-\mathrm{PL_B})/\mathrm{PL_B}$. We observed non-monotonic behavior $C (\tau)$, reflecting Rabi oscillations that decay rapidly. This rapid decay likely arises from several closely spaced resonances associated with the hyperfine structure, which lead to multiple oscillation frequencies. By fitting the rapidly decaying Rabi oscillations, we determine the RF pulse duration corresponding to a $\pi$ rotation on the Bloch sphere and use this value for the $T_1$ measurements following the protocol shown in Fig.~\ref{fig4}a.

A mono-exponential fit $\exp (- \tau/ T_1)$ to the data in Fig.~\ref{fig4}b yields $T_1 = 660 \, \pm \, 40 \,\mathrm{ns}$. This value is many orders of magnitude shorter than expected for phonon-mediated spin-lattice relaxation of spin centers in SiC at low temperature \cite{10.1103/physrevb.95.161201}. We therefore assume that the measured relaxation time is limited by the metastability of the charge state. One possible scenario is that optical excitation induces charge conversion from the neutral state $\mathrm{ClV^0}$ to the spin-active positively charged state $\mathrm{ClV^+}$ by promoting an electron from the defect into the conduction band. This process is analogous to the photo-induced ionization dynamics of the $\mathrm{NV^-}$/$\mathrm{NV^0}$ defects in diamond \cite{10.1088/1367-2630/15/1/013064}. After the laser excitation is switched off, the $\mathrm{ClV^+}$ state relaxes back to the neutral $\mathrm{ClV^0}$ state, leading to quenching of the ODMR contrast on this timescale. The ODMR contrast does not fully quench but saturates at about $6 \%$ of its maximum value. This suggests that a fraction of defects remains in the $\mathrm{ClV^+}$ state in the dark and may contribute to the observed coherent spin dynamics. Further experiments will be required to clarify this mechanism and are beyond the scope of the present work. We emphasize that the observed spin lifetime is nevertheless significantly longer than the ES lifetime extracted from the PL decay measurements of Fig.~\ref{fig2}c. This provides strong evidence that the ODMR signal does not originate from the ES spins. 

To realize coherent driving of the electron spins at the $\nu_1$ resonance, we employ a standard Ramsey protocol to measure the $T_2^*$ time (Fig.~\ref{fig4}a). Following optical initialization into the $m_S = 0$ state, a first RF $\pi/2$ pulse creates a coherent superposition of the $m_S = 0$ and $m_S = \pm 1$ states, which evolves during a free precession time $\tau$. A second RF $\pi/2$ pulse then converts the accumulated phase into a population difference, which is read out optically via the spin-dependent PL. We observe an exponential decay with a characteristic time of  $660  \,\mathrm{ns}$, indicating that the spin coherence time $T_2^*$, similar to the spin relaxation time $T_1$, is limited by quenching of the ODMR contrast. After subtracting the quenching background, clear Ramsey fringes are revealed in Fig.~\ref{fig4}c. 

The fast Fourier transform (FFT) spectrum of the Ramsey fringes reveals two dominant oscillation frequencies $\omega_1$ and $\omega_2$. These frequencies correspond to the detuning of the driving RF frequency from the transition between the spin states $|\mp 1, \pm 1/2\rangle$  and $|0, \pm 1/2\rangle$ for $\omega_1$ and $|\mp 1, \pm 3/2\rangle$ and  $|0, \pm 3/2\rangle$ for $\omega_2$. Because these transitions are closely spaced in frequency, they are unresolved in the ODMR spectrum in Fig.~\ref{fig1}c and merge into a single transition $\nu_1$ in the scheme of Fig.~\ref{fig3}b. We emphasize that even a weak residual magnetic field lifts the degeneracy of these states and induces mixing, resulting in up to 16 closely spaced spin transitions and a more complex spectrum in the frequency domain and, correspondingly,  complex Ramsey pattern in the time domain. From the best fit of the experimental data in Fig.~\ref{fig4}c to 
\begin{align}\label{eq:Ramsey}
C (\tau) = (a_1 \cos \omega_1 \tau + a_2 \cos \omega_2 \tau) \, e^{- \tau / T_2^* }  
\end{align}
with $\omega_1 / 2 \pi = 1.6 \, \mathrm{MHz}$ and $\omega_2 / 2 \pi = 10.1 \, \mathrm{MHz}$, we estimate  the intrinsic inhomogeneous spin coherence time $T_2^* = 2 \pm 1 \, \mathrm{\mu s}$, which is comparable with that for ensemble of intinsic defects in SiC \cite{10.1103/physrevb.95.161201}.

\section{Outlook}

Our experiments establish chlorine-related defects in SiC as a promising platform for telecom-band quantum technologies. The qualitative agreement with theoretical predictions supports the identification of the underlying defect configurations, while remaining quantitative discrepancies indicate that current models do not yet fully capture the electronic structure and spin properties. In particular, the unexpectedly large Debye-Waller factor and the involvement of additional excited or charge states in the optical dynamics point to a more complex behaviour than previously anticipated. These findings call for refined theoretical descriptions of the ClV center and related chlorine-based defect configurations. Complementary experimental studies will be required to achieve a comprehensive microscopic understanding of this defect family.

Looking forward, the isolation of single chlorine-based emitters represents a critical step toward their deployment in quantum photonic architectures. Their integration into nanophotonic cavities and electrically active devices will enable enhanced photon extraction rates and scalable, semiconductor-compatible platforms. Key challenges include controlling the charge state through doping and photoexcitation, as well as achieving high spectral stability under resonant excitation and device operation conditions. At the same time, clarifying the optical spin polarization cycle will be essential for establishing robust spin initialization and readout. Extending experiments to higher magnetic fields will further enable advanced spin control, including the manipulation of coupled nuclear spins. This opens a pathway toward multi-qubit quantum registers interfaced with telecom photons.

In conclusion, chlorine-related defects in SiC emerge as a compelling platform for integrated quantum technologies that combine telecom-band operation with scalable device integration. While important questions regarding the ISC process and electronic structure remain, a progress in single-emitter isolation, device integration and advanced spin control is expected to establish this defect system as a versatile building block for wafer-scale quantum photonic networks.

During the preparation of this manuscript, we became aware of an independent experimental study observing similar telecom-band ZPLs for Cl-related defects in 4H-SiC, consistent with our earlier seminal work \cite{10.1364/oe.581705}, but employing a reversed labeling convention and a different assignment to possible configurations of Cl-related defects \cite{10.48550/arxiv.2602.08854}. That study reports ODMR resonances in the frequency range of $1.0 - 1.4 \, \mathrm{GHz}$, which are not identified in our experiments. In addition, no time-resolved measurements or coherent spin dynamics are performed. These observations may be consistent with the presence of multiple configurations of Cl-related defects in SiC.

\vspace{1em}

%\section*{Data availability}
%The data are available upon request. 

\section*{Acknowledgments}
%\begin{acknowledgments}
This work was funded by the European Union under project 101186889 QuSPARC and the Leibniz Association via grant No. K741/2025 TeleQuaM. Support from the Ion Beam Center (IBC) at HZDR for ion implantation is gratefully acknowledged. 

%\end{acknowledgments}

%\section*{Author contributions}
%To be written later... 

%\section*{Competing interests}
%The authors declare no competing interest. 

%***********************************
%\bibliography{ClV_references} 
%***********************************

%

\end{document}